\documentclass[aps]{revtex4-1}
\usepackage{amsmath}
\usepackage{amsthm}
\usepackage{graphicx}
\usepackage{multirow}
\usepackage{longtable}
\begin{document}
\title{Exact test for Markov order}
\author{Shawn D. Pethel and Daniel W. Hahs*}
\affiliation{U.S. Army RDECOM, RDMR-WSS, Redstone Arsenal, Alabama
35898, USA; *Torch Technologies, Inc., Huntsville, AL 35802}
\date{\today}
\begin{abstract}
We describe an exact test of the null hypothesis that a Markov chain
is nth order versus the alternate hypothesis that it is $(n+1)$-th
order. The procedure does not rely on asymptotic properties, but
instead builds up the test statistic distribution via surrogate data
and is valid for any sample size. Surrogate data are generated using
a novel algorithm that guarantees, per shot, a uniform sampling from
the set of sequences that exactly match the nth order properties of
the observed data.
\end{abstract}
\maketitle
\par
It often happens that it is useful to describe a process as a set of
discrete states with probabilistic transitions. Examples abound in
various fields such as the study of chemical processes \cite{Tamir},
DNA sequences \cite{Avery}, finance \cite{Jarrow}, and nonlinear
dynamics \cite{Hao}, among others. If a transition to a new state is
conditioned only on the present state we call this model a Markov
chain. An $n$th-order Markov chain is a generalization to include
the past $n$ states in the transition probability. When the
conditional probabilities are not otherwise given, they are
estimated from a time series of observations.
\par
If the order of the Markov chain is in question there are various
tests and criteria available to narrow down the options. A classical
approach is to formulate the question as a hypothesis test that a
chain is $n$-th order versus $(n+1)$-th order\cite{Anderson}. When
the test statistic has a known limiting distribution, such as
$\chi^2$, a $p$-value can be calculated and a decision made based on
a chosen significance level\cite{Greenwood}. Another avenue are the
information criteria tests such as AIC and BIC
\cite{Tong,Katz,Burnham}. These produce rankings over multiple
orders based on expected likelihood and have built-in corrections
for over-fitting. Both of these approaches rely on approximations
that are only valid in the limit of large samples. In small sample
situations one cannot be sure of their efficacy.
\par
It is possible to perform an exact hypothesis test that is valid for
any sample size. Instead of relying on asymptotic properties, the
test statistic distribution is discovered by generating samples
(referred to here as \emph{surrogates}) that exactly match the
$n$-th order properties of the observed time series \cite{Heyden}.
The challenge is in efficiently generating a large number of such
samples, especially for higher orders. To our knowledge no solution
to this problem has been reported in the literature. The
contribution of this work is a surrogate data procedure that has
ideal properties: one sample is generated per shot, samples are
uniformly selected from the set of all possible surrogate sequences,
computation time increases linearly with the length of the sequence,
and any order can be accommodated. Armed with this new procedure it
is now practical to perform exact hypothesis tests of Markov order.
\par
We first describe how to do hypothesis testing of Markov order using
the $\chi^2$ statistic, for which the distribution is known in the
large sample limit. Next we describe the method of surrogate data
generation based on Whittle's formula. Then we compare the $\chi^2$
statistic in large and small sample cases using both the asymptotic
distribution and the exact distribution obtained from the
surrogates.
\par
A sequence of observations $\{x_1\ldots x_N\}$ form a Markov chain of
order $n$ if the conditional probability satisfies
\begin{equation}
p(x_{t+1}|x_{t},x_{t-1}\ldots)=p(x_{t+1}|x_{t}\ldots x_{t-n+1}),
\label{markovorder}
\end{equation}
for all $n<t\le N$. For convenience we will label the states each
measurement can take by positive integers up to $M$. A sequence of
discrete measurements may come from a process that is naturally
discrete, such as a DNA sequence, or from a continuous process that
has been discretized by an analog-to-digital measuring device.
Unless otherwise specified a Markov process is assumed to be first
order $(n=1)$. This means that the transition probabilities to a
future state depend only on the present state and not on prior
states. An $n$th order process can always be cast as first order by
grouping the present state with the relevant past states into a
\emph{word}, in which case the number of states can be up to $M^n$.
A process that has no dependence on past or present (such as a
random iid process) is said to be zeroth order.
\par
To perform a hypothesis test of $n$-th order versus $(n+1)$-th order
one begins with an assumption of $n$th order and then computes the
distribution of a suitable $(n+1)$-th order statistic. If the
observed $(n+1)$-th order statistic is highly unlikely then $n$-th
order assumption is rejected. The initial assumption is called the
null hypothesis and the probability of the observed statistic given
the distribution implied by the null hypothesis is referred to as
the p-value. Typically, a p-value less than or equal to $0.05$ is
taken as grounds to reject the null hypothesis.
\par
Let us begin with the assumption that $\{x_t\}$ is first order
($n=1$) and calculate the p-value of a second order statistic using
a $\chi^2$ distribution. The null hypothesis is
\begin{equation}
p(x_{t+1}=i|x_t=j,x_{t-1}=k) = p(x_{t+1}=i|x_t=j),
\end{equation}
or using Bayes' rule
\begin{equation}
p(x_{t+1}\!=i,x_t\!=j,x_{t-1}\!=k) =
\frac{p(x_{t+1}\!=i,x_t\!=j)p(x_t\!=j,x_{t-1}\!=k)}{p(x_t\!=j)}.
\label{mtest}
\end{equation}
The l.h.s. of (\ref{mtest}) multiplied by $N-2$ is the expected
number of times the word $(x_{t+1}\!=\!i,x_t\!=\!j,x_{t-1}\!=\!k)$
appears in the data given the null hypothesis. The quantities on the
r.h.s. are not expected values; they are taken from the observed
sequence. Let $E_w$ be the expected word count where $\sum E_w=N-2$
and $w$ indexes the set of all words for which the expected count is
greater than zero. Similarly, let $O_w$ be the corresponding count
from the observed data. If the sequence $w^\prime$ does not appear
in the observed data, then $O_{w^\prime}=0$. We can now define the
expected $\chi^2$ statistic
\begin{equation}
\chi_{\text{exp}}^2=\sum_w\frac{(E_w-O_w)^2}{E_w},
\end{equation}
which is a measure of the deviation of the observed count from the
expected. The first order assumption does not uniquely determine the
second order statistics; there is some freedom for $\chi^2$ to vary from
shot to shot even assuming the null hypothesis is true. The advantage of
the $\chi^2$ statistic is that, given the degrees of freedom $d$, the
distribution $f(\chi^2;d)$ is known in the limit
$N\rightarrow\infty$. The p-value is then obtained by integrating
$f(\chi^2)$ over $\chi^2\ge\chi_\text{exp}^2$.
\par
A detail not made clear in the literature is how to compute the
$(n+1)$-th order degrees of freedom $d$ needed to determine the
$\chi^2$ distribution. Let $F$ be a transition count matrix of size
up to $M^n\times M^n$. The $(i,j)$-th entry of $F$ is the number of
times word $i$ transitions to word $j$. Because words overlap and
differ by only one observation, there are at most $M$ nonzero
entries in row $i$ of $F$. In the case that all words are present,
$F$ can be rearranged in block diagonal form with $n$ $M\times M$
blocks. In the case that some words are not present in the observed
data, these blocks will be of differing size. If the size of the
$k$th block is $r_k\times c_k$, then the total number of degrees of
freedom $d$ is $\sum (r_k-1)(c_k-1)$.
\par
The hypothesis test as described above is not exact; it relies on
the $\chi^2$ distribution in the asymptotic limit of infinite data.
To discover the exact distribution for finite data one needs to know
all possible sequences that satisfy the null hypothesis along with
their likelihood. For the first order hypothesis these sequences
must all have exactly the same joint probabilities shown on the
r.h.s. of (\ref{mtest}). Let $F_{ij}$ be the $(x_{t+1}\!=\!i,x_t\!=\!j)$
word count for the observed sequence and let $S$ represent the set
of sequences with the same $F$ and the same beginning and end state
as the observed sequence. The members of $S$ all have the same
$\chi_\text{exp}^2$ as the observed sequence (but not necessarily
the same second order statistics).
\par
The number of sequences that have the word count $F$ and begin with
state $u$ and end with state $v$ is given by Whittle's formula
\cite{Billingsley}:
\begin{equation}
N_{uv}(F) = \frac{\Pi_i F_{i\cdot}!} {\Pi_{ij} F_{ij}!} C_{vu}
\label{whittle}
\end{equation}
where $F_{i\cdot}$ is the sum of row $i$ and $C_{vu}$ is the
($v,u$)-th cofactor of the matrix
\begin{equation}
F^*_{ij}=
\begin{cases}
\delta_{ij}-F_{ij}/F_{i\cdot} & \text{if $F_{i\cdot}>0$},\\
\delta_{ij} & \text{if $F_{i\cdot}=0$.}
\end{cases}
\label{minor}
\end{equation}
The cofactor is computed by striking out the $v$-th row and $u$-th
column and taking the determinant.
\par
As an example, consider the following sequence of twelve
binary observations:
\begin{equation}
\mathbf{x} = \{0\ 1\ 1\ 0\ 1\ 0\ 1\ 1\ 1\ 0\ 0\ 1\}.
\end{equation}
The sequence $\mathbf{x}$ has $u=0$, $v=1$ and transition count
\begin{equation}
F =
\begin{pmatrix} 1 & 4 \\ 3 & 3\end{pmatrix}.
\end{equation}
From (\ref{minor}) we compute
\begin{equation}
F^* =
\begin{pmatrix} \frac{4}{5} & -\frac{4}{5} \\ -\frac{1}{2} & \frac{1}{2}\end{pmatrix}
\end{equation}
and $C_{10}=\text{det}(4/5) = 4/5$. Plugging into (\ref{whittle})
gives
\begin{equation}
N_{01}(F)=\frac{5!\cdot 6!}{3!\cdot 3!\cdot 4!}\cdot \frac{4}{5}=80.
\end{equation}
The cardinality of the set $S(\mathbf{x})$ is therefore $80$. The
transition count $F$ determines the first order joint probabilities
$p(x_{n+1},x_n)$ and, after fixing the first and the last symbol,
the zeroth order probabilities $p(x_n)$ as well. Therefore all $80$
sequences in $S$ have first order transition probabilities
$p(x_{n+1}|x_n)$ identical to the observed sequence $\mathbf{x}$.
\par
For all but the shortest sequences the value of (\ref{whittle}) is
so large that it cannot be computed using fixed precision
arithmetic. In our algorithm we instead compute the natural
logarithm of (\ref{whittle}) using a Stirling's series for the
factorial terms:
\begin{equation}
\ln z! \sim z\ln z - z + \frac{1}{2}\ln(2\pi
z)+\frac{1}{12z}-\frac{1}{360z^3}+\frac{1}{1260z^5}-\frac{1}{1680z^7}
\end{equation}
when $z > 16$.
\par
To find the p-value we need to know the fraction of sequences have
$\chi^2$ values less than or equal to $\chi_\text{exp}^2$. If $|S|$
is too large to enumerate all the sequences, the $p$-value can still
be estimated to any desired accuracy provided one has a method of
producing uniform random samples from the set $S$. Previously
reported methods for generating samples from $S$ are impractical,
especially for higher order testing \cite{Heyden}. Here we give an
efficient procedure.
\begin{figure}
\includegraphics[scale=0.55]{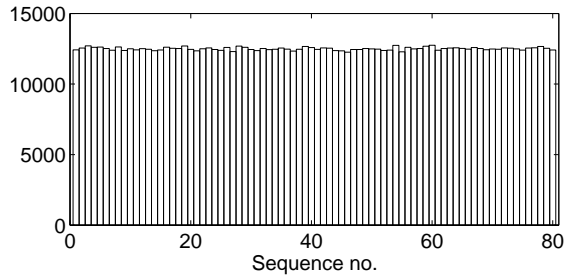}
\caption{The number of times a sequence appears in $10^6$ iterations
of Whittle's algorithm. The sequences labeled $1-80$ refer to the
example in the text.} \label{figure1}
\end{figure}
\par
We construct a member $\{y_t\}$ of $S$ starting with $y_1=u$,
ending with $y_N=v$, and having the transition count matrix $F$. The candidates for
the second element $y_2$ are the set $\{y_2|F_{y_1y_2}>0\}$. For each candidate we
compute $N_{y_2v}(F^\prime)$, the number of sequences left, where
$F_{ij}^\prime=F_{ij} -\delta_{y_1y_2}$. We choose a candidate
randomly in proportion to the number of sequences left; a path that
leads to a small number of possibilities is chosen less frequently
than one that leads to a large number. Once $y_2$ is chosen $F$ is
set equal to the appropriate $F^\prime$ and the process is repeated
for $y_3$ and so on until $y_{N-1}$ is reached.
\par
Returning to our example case, we have $y_1=0$, $y_{12}=1$, and
$y_2=\{0,1\}$. The two choices for $y_2$ lead to the following
number of sequences:
\begin{eqnarray}
N_{01}\begin{pmatrix} 0 & 4 \\ 3 & 3\end{pmatrix} & =& 20,\nonumber\\
N_{11}\begin{pmatrix} 1 & 3 \\ 3 & 3\end{pmatrix} & =& 60.
\end{eqnarray}
Therefore $y_2=0$ is chosen with $20/80=1/4$ probability and $y_2=1$
with $3/4$ probability. By weighting our choice at each step by
Whittle's formula we guarantee paths that do not result in a valid
sequence are not followed and that all valid paths are followed with
uniform probability (Fig.\ref{figure1}). This method is suitable for
the generation of very long surrogates, as the difficulty increases
only linearly with $N$. For producing sequences of order $n>1$
simply replace the elements of ${y_t}$ with length $n$ words. The
matrix $F$ can be as large as $M^n\times M^n$, but has no more than
$N-n$ nonzero elements and can be handled efficiently using sparse
methods. Code is available for generating surrogates by this method \cite{Pethel}.
\begin{figure}
\includegraphics[scale=0.45]{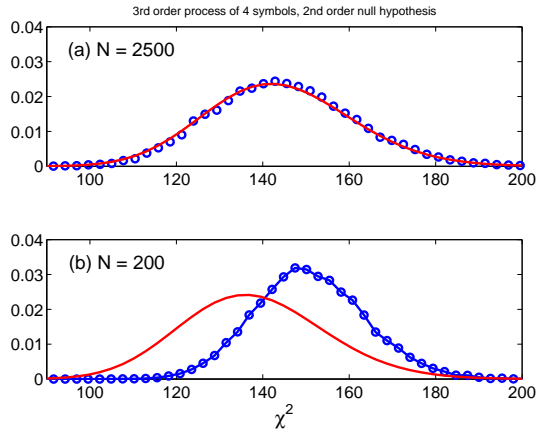}
\caption{The asymptotic $\chi^2$ probability density (no marker) and
the actual density estimated using surrogate data (circles). Data is
taken from a randomly generated Markov process, degrees of freedom
$d$ are computed for $2$nd order statistics given a $1$st order null
hypothesis. Top panel (a) utilizes $2500$ data points whereas the
bottom panel (b) only $200$. For the short time series (b) the
asymptotic distribution differs considerably from the actual. }
\label{figure2}
\end{figure}

\par
Figure (\ref{figure2}) shows the $\chi^2$ density computed in the
asymptotic limit with the density estimated from $20000$ surrogates
of a random Markov process of 4 states. The degrees of freedom are
calculated for $2$nd order assuming fixed $1$st order statistics. In
the top panel ($N=2500$) there is close agreement indicating that
that the surrogate data statistics behave as expected in the
asymptotic limit. The bottom panel uses the same time series, but
only the first $200$ data points.  The significant disagreement
between the two densities illustrates the need for an exact test
when the sample is small.
\par
The efficacy of a hypothesis test is quantified by its \emph{size}
and \emph{power}. The size of a test is its probability of
incorrectly rejecting the null hypothesis (Type I error). For an
ideal test the size should be equal to the significance level
($0.05$). The power of a test is its probability of \emph{correctly}
rejecting the null hypothesis. The failure to do so is a Type II
error. To estimate power we use data from Markov processes that are
one order higher than the null hypothesis; other choices could yield
different results.
\par
Test cases are taken from a set of randomly generated $n$th order
Markov processes with four states ($M=4$). Recall that such a
process is specified by the transition probabilities
$p(x_{t+1}|x_{t}\ldots x_{t-n+1})$, which when expressed as a matrix
is size $M^n\times M$. One way to create such a transition matrix is
to populate it with $[0,1]$ random numbers and normalize the rows.
We have found, however that this procedure tends to produce weakly
$n$th order processes, particularly when either or both $M$ and $n$
are large. To produce strongly $n$th order processes we first scale
the random numbers by adding one, raising them to the $10$th power,
and then row normalizing. This creates more variance in the
transition probabilities.
\par
We generated $2500$ trials for each size and power estimate shown in
the following tables. For each case we tabulate results using the
asymptotic distribution (labeled $\chi^2$) and the exact $\chi^2$
distribution (labeled
$\chi^2_{\text{surg}}$) obtained using $2500$ surrogates. In addition we show $H_{\text{surg}}$,
which relies on the same surrogate data, but instead of the $\chi^2$ statistic the
$n$th order entropy rate is used:
\begin{equation}
H(x_{t+1}|x_{t}\ldots x_{t-n+1})= 
H(x_{t+1},x_{t}\ldots x_{t-n+1})-H(x_{t}\ldots x_{t-n+1}).
\end{equation}
As the surrogates have identical $n$th order block entropies, only
$H(x_{t+1},x_{t}\ldots x_{t-n+1})$ needs to be re-computed for each
trial. The use of this statistic was suggested in \cite{Heyden}.

\begin{table}[h]
 \setlength{\tabcolsep}{8pt}
\begin{tabular}{lccclcccl}
\hline
 & & \multicolumn{7}{c}{1st Order}\\
 & & \multicolumn{3}{c}{Size $\pm 0.01$} &  &\multicolumn{3}{c}{Power $\pm 0.01$}\\
\cline{3-9} Data & & $\chi^2$ & $\chi^2_{\text{surg}}$\ &
$H_{\text{surg}}$& & $\chi^2$ & $\chi^2_{\text{surg}}$ &
$H_{\text{surg}}$\\
\hline
$25$ & & $0.04$ & $0.04$ & $0.03$ & & $0.48$ & $0.49$ & $0.49$\\
$50$ & & $0.05$ & $0.05$ & $0.04$ & & $0.89$ & $0.89$ & $0.91$\\
$100$ & & $0.06$ & $0.04$ & $0.05$ & & $0.98$ & $0.98$ & $0.98$\\
$200$ & & $0.08$ & $0.05$ & $0.05$ & & $1.00$ & $1.00$ & $1.00$\\
$400$ & & $0.09$ & $0.05$ & $0.05$ & & $1.00$ & $1.00$ & $1.00$\\
\hline
\end{tabular}
 \caption{\label{tab1}Estimated size of asymptotic and exact
$\chi^2$ statistic for random $1$st order Markov processes with $4$
symbols, $2500$ trials. }
\end{table}

\begin{table}[h]
\setlength{\tabcolsep}{8pt}
\begin{tabular}{lccclcccl}
\hline
 & & \multicolumn{7}{c}{2nd Order}\\

 & & \multicolumn{3}{c}{Size $\pm 0.01$} &  &\multicolumn{3}{c}{Power $\pm 0.01$}\\
\cline{3-9} Data & & $\chi^2$ & $\chi^2_{\text{surg}}$\ &
$H_{\text{surg}}$& & $\chi^2$ & $\chi^2_{\text{surg}}$ &
$H_{\text{surg}}$\\
\hline
$50$ & & $0.07$ & $0.05$ & $0.04$ & & $0.79$ & $0.59$ & $0.56$\\
$100$ & & $0.10$ & $0.05$ & $0.05$ & & $0.98$ & $0.96$ & $0.98$\\
$200$ & & $0.10$ & $0.05$ & $0.05$ & & $1.00$ & $1.00$ & $1.00$\\
$400$ & & $0.11$ & $0.05$ & $0.06$ & & $1.00$ & $1.00$ & $1.00$\\
\hline
\end{tabular}
\caption{\label{tab2}Estimated size of asymptotic and exact $\chi^2$
statistic for random $2$nd order Markov processes with $4$ symbols,
$2500$ trials. }
\end{table}

\begin{table}[h]
\setlength{\tabcolsep}{8pt}
\begin{tabular}{lccclcccl}
\hline
 & & \multicolumn{7}{c}{3rd Order}\\

 & & \multicolumn{3}{c}{Size $\pm 0.01$} &  &\multicolumn{3}{c}{Power $\pm 0.01$}\\
\cline{3-9} Data & & $\chi^2$ & $\chi^2_{\text{surg}}$\ &
$H_{\text{surg}}$& & $\chi^2$ & $\chi^2_{\text{surg}}$ &
$H_{\text{surg}}$\\
\hline
$50$ & & $0.07$ & $0.01$ & $0.01$ & & $0.44$ & $0.04$ & $0.03$\\
$100$ & & $0.18$ & $0.05$ & $0.05$ & & $0.97$ & $0.56$ & $0.52$\\
$200$ & & $0.22$ & $0.05$ & $0.05$ & & $1.00$ & $0.99$ & $0.99$\\
$400$ & & $0.22$ & $0.05$ & $0.05$ & & $1.00$ & $1.00$ & $1.00$\\
\hline
\end{tabular}
\caption{\label{tab3}Estimated size of asymptotic and exact $\chi^2$
statistic for random $3$rd order Markov processes with $4$ symbols,
$2500$ trials. }
\end{table}

\begin{table}[h]
\setlength{\tabcolsep}{8pt}
\begin{tabular}{lccclcccl}
\hline
 & & \multicolumn{7}{c}{4th Order}\\

 & & \multicolumn{3}{c}{Size $\pm 0.01$} &  &\multicolumn{3}{c}{Power $\pm 0.01$}\\
\cline{3-9} Data & & $\chi^2$ & $\chi^2_{\text{surg}}$\ &
$H_{\text{surg}}$& & $\chi^2$ & $\chi^2_{\text{surg}}$ &
$H_{\text{surg}}$\\
\hline
$50$ & & $0.01$ & $0.00$ & $0.00$ & & $0.02$ & $0.00$ & $0.00$\\
$100$ & & $0.12$ & $0.01$ & $0.01$ & & $0.61$ & $0.03$ & $0.02$\\
$200$ & & $0.49$ & $0.04$ & $0.03$ & & $1.00$ & $0.44$ & $0.42$\\
$400$ & & $0.74$ & $0.05$ & $0.05$ & & $1.00$ & $0.99$ & $0.99$\\
$800$ & & $0.77$ & $0.05$ & $0.05$ & & $1.00$ & $1.00$ & $1.00$\\
\hline
\end{tabular}
\caption{\label{tab4}Estimated size of asymptotic and exact $\chi^2$
statistic for random $4$th order Markov processes with $4$ symbols,
$2500$ trials.}
\end{table}
\par
We break out each order in a separate table and list size and power
versus data length. In the large sample limit both the exact and
asymptotic methods should approach a power of $1$ and a size equal
to the significance level ($0.05$). The exact test is quite
efficient; as little as $100$ data points are needed for $1$st and
$2$nd order tests, $200$ for $3$rd order, and $400$ for $4$th order.
The asymptotic method is very slow to attain the ideal size even for
the $1$st order test ($10^6$ sample size, not shown). For higher
order tests we do not recommend use of the $\chi^2$ distribution.
There is no detectable difference between using the entropy rate as
a test statistic and $\chi^2$. As entropy rate is a simpler quantity
to calculate, we recommend its use over $\chi^2$.
\par
Compared to the asymptotic $\chi^2$ test, the exact test is much
slower; each step of Whittle's algorithm requires the computation of the
determinate of the transition matrix. Considering that the transition matrix
changes in only one entry at each step there is potential to improve the
efficiency over our naive implementation. Even without such optimizations,
it is well within a desktop computer's ability to
generate many thousands of surrogate sequences in minutes. Because
each surrogate is generated independently, parallelization is
straightforward. For our tables, each case involving $2500$ trials,
we opted to use $2500$ surrogates, requiring the generation of $6.25$ million surrogates per table entry. 
The standard error of our
$p$-value estimates as well as our size and power estimates is then
$1/\sqrt{4\times 2500}$ or $\pm 0.01$. If one doesn't need to analyze so
many data sets, we recommend using $10,000$ or more surrogates. 
\par
In summary, we have described an exact test of the null hypothesis
that a Markov chain is nth order versus the alternate hypothesis
that it is $(n+1)$-th order. At the heart of the test is an
algorithm based on Whittle's formula, which efficiently produces
surrogate data sets that have identical word transition counts as
the observed sequence. Whittle's algorithm together with the entropy
rate statistic make for a conceptually simple approach to Markov
order hypothesis testing; no calculation of degrees of freedom or
corrections for small sample size are necessary.

\end{document}